\documentclass{aastex61}
\usepackage{amsmath,CJK}

\received{--}
\revised{--}
\accepted{--}

\shorttitle{The Fate of Lexell's Comet}
\shortauthors{Ye, Wiegert \& Hui}

\begin{document}
\begin{CJK*}{UTF8}{gbsn}

\title{Finding Long Lost Lexell's Comet: The Fate of the First Discovered Near-Earth Object}

\correspondingauthor{Quan-Zhi Ye}
\email{qye@caltech.edu}

\author[0000-0002-4838-7676]{Quan-Zhi Ye (叶泉志)}
\affiliation{Division of Physics, Mathematics and Astronomy, California Institute of Technology, Pasadena, CA 91125, U.S.A.}
\affiliation{Infrared Processing and Analysis Center, California Institute of Technology, Pasadena, CA 91125, U.S.A.}

\author{Paul A. Wiegert}
\affiliation{Department of Physics and Astronomy, The University of Western Ontario, London, Ontario N6A 3K7, Canada}
\affiliation{Centre for Planetary Science and Exploration, The University of Western Ontario, London, Ontario N6A 5B8, Canada}

\author{Man-To Hui (许文韬)}
\affiliation{Department of Earth, Planetary and Space Sciences, UCLA, Los Angeles, CA 90095-1567, USA}



\begin{abstract}
Jupiter-family Comet D/1770 L1 (Lexell) was the first discovered Near-Earth Object (NEO), and passed the Earth on 1770 Jul 1 at a recorded distance of 0.015~au. The comet was subsequently lost due to unfavorable observing circumstances during its next apparition followed by a close encounter with Jupiter in 1779. Since then, the fate of D/Lexell has attracted interest from the scientific community, and now we revisit this long-standing question. We investigate the dynamical evolution of D/Lexell based on a set of orbits recalculated using the observations made by Charles Messier, the comet's discoverer, and find that there is a $98\%$ chance that D/Lexell remains in the Solar System by the year of 2000. This finding remains valid even if a moderate non-gravitational effect is imposed. Messier's observations also suggest that the comet is one of the largest known near-Earth comets, with a nucleus of $\gtrsim 10$~km in diameter. This implies that the comet should have been detected by contemporary NEO surveys regardless of its activity level if it has remained in the inner Solar System. We identify asteroid 2010 JL$_{33}$ as a possible descendant of D/Lexell, with a $0.8\%$ probability of chance alignment, but a direct orbital linkage of the two bodies has not been successfully accomplished. We also use the recalculated orbit to investigate the meteors potentially originating from D/Lexell. While no associated meteors have been unambiguously detected, we show that meteor observations can be used to better constrain the orbit of D/Lexell despite the comet being long lost.
\end{abstract}




\section{Introduction} \label{sec:intro}

Jupiter-family Comet D/1770 L1 (Lexell) was the first known Near-Earth Object (NEO)\footnote{The first observed NEO known to date is Comet 1P/Halley, recorded by Chinese chroniclers in 240~BC \citep[e.g.][]{Stephenson1985}.}. Found by Charles Messier \citep{Messier1776} and named after its orbit computer Anders Johan Lexell \citep{Lexell1778a}, D/Lexell approached to a distance of only 0.015~au from the Earth on 1770 Jul 1, a record that has not been surpassed by any known comet so far\footnote{Comets C/1491 B1 and P/1999 J6 (SOHO) may have passed closer than D/Lexell at their respective close approach to the Earth in 1491 and 1999, but their orbits are somewhat uncertain, therefore the approach distance of each comet cannot be precisely calculated.}. Although the orbit calculated by Lexell showed a period of 5.58~yrs, the comet was not seen after 1770. In his celebrated work, \citet{Lexell1778a} suggested that a close approach to Jupiter in 1779 had perturbed the comet into a high perihelion orbit, while the comet was behind the Sun as seen from the Earth during its 1776 perihelion and was therefore unobservable. This result was confirmed by Johann Karl Burckhardt \citep{Burckhardt1807}, winning him a prize dedicated to this problem offered by the Paris Academy of Sciences. The work by \citet{Leverrier1844a,Leverrier1844b} reconfirmed the results by Lexell and Burckhardt and provided a very complete review of the matter.

Despite the consensus that D/Lexell has evolved into a very different orbit, the interest about the fate of the comet is long-lived. Some 80 years later, \citet{Chandler1889,Chandler1890} suggested that the newly-discovered 16P/Brooks could be the return of D/Lexell. It took another 15 years for \citet{Poor1905} to demonstrate that such linkage was unlikely. After the 1950s, the development of meteor astronomy sparked searches for meteor activity associated with D/Lexell \citep{Nilsson1963,Kresakova1980,Carusi1982,Carusi1983,Olsson-Steel1988}, although no definite conclusions have been reached.

The recent years have witnessed tremendous progress in the studies of NEOs and their dust production. We have reached 90\% completion of NEOs greater than 1~km in diameter \citep{Jedicke2015}. Some 800 meteoroid streams have been reported by various radio and video meteor surveys \citep{Jenniskens2017}, many without an identified parent NEO. These new data would benefit a renewed search for D/Lexell and/or its descendants. Here we present a reexamination of topic using the original observations of D/Lexell and the most recent observations of NEOs and meteor showers.

\section{Reconstruction of Orbit} \label{sec:orbit}

Almost all of the surviving astrometric measurements of D/Lexell were made by Messier, who observed the comet from his discovery of it on 1770 Jun 14 through Oct 3,  when he was also the last astronomer to observe the comet. Since we have no reason to believe that the few other measurements would be of significantly higher quality than Messier's, we focus exclusively on Messier's observations, which are available from \textit{Memoires~de~l'Academie~Royale~des~Sciences} \citep{Messier1776}. These observations were taken in 18th century Paris, so they referred to the Paris meridian, which is $2^\circ20'14''$ east of the now-used Greenwich meridian. The astronomical time in the 18th century also started at noon. We correct for the different meridian and time definitions and assume the positions Messier reports refer to the epoch of the observation, which we precessed to the J2000 epoch. The corrected positions are tabulated in Table~\ref{tbl:obs}.

\startlongtable
\begin{deluxetable}{llll}
\tablecaption{Messier's observations of D/Lexell, precessed to J2000 epoch. All observations were taken at Paris (Minor Planet Center Observatory Code 007)} \label{tbl:obs}
\tablehead{
\colhead{Time (UT)} & \colhead{R.A.} & \colhead{Dec.} & \colhead{Note} \\
}
\startdata
1770 Jun 14.97256 & 18h24m52.9s & -16$^\circ$39$'$54$''$ & \\
1770 Jun 15.96809 & 18 25 06.7 & -16 22 54 & \\
1770 Jun 17.95989 & 18 25 32.7 & -15 40 59 & \\
1770 Jun 20.93853 & 18 26 22.7 & -14 13 33 & \\
1770 Jun 21.92948 & 18 26 44.0 & -13 34 18 & \\
1770 Jun 22.92798 & 18 27 11.2 & -12 43 16 & \\
1770 Jun 23.00022 & 18 27 14.8 & -12 39 34 & \\
1770 Jun 26.05458 & 18 29 33.1 & -08 22 30 & \\
1770 Jun 28.04442 & 18 32 46.5 & -02 04 19 & \\
1770 Jun 28.94253 & 18 35 42.7 & +03 20 00 & \\
1770 Jun 29.99314 & 18 42 02.1 & +14 57 42 & \\
1770 Jun 30.99575 & 18 58 35.1 & +37 48 06 & estimated without instrument during break in clouds \\
1770 Jul 01.99588 & 21 34 06.1 & +78 01 52 & estimated without instrument \\
1770 Jul 03.95447 & 06 13 37.5 & +48 58 23 & estimated without instrument while at Minister of State's house \\
1770 Aug 03.12079 & 06 39 59.4 & +22 18 30 & \\
1770 Aug 04.10822 & 06 41 36.5 & +22 13 39 & \\
1770 Aug 05.08575 & 06 43 11.3 & +22 09 53 & \\
1770 Aug 06.10375 & 06 45 04.7 & +22 04 34 & \\
1770 Aug 07.09749 & 06 46 52.0 & +22 00 36 & comet viewed with difficulty, observations doubtful \\
1770 Aug 08.11111 & 06 48 50.5 & +21 55 37 & \\
1770 Aug 09.09021 & 06 50 43.5 & +21 50 28 & \\
1770 Aug 10.11031 & 06 52 55.4 & +21 44 45 & \\
1770 Aug 11.08690 & 06 54 56.0 & +21 42 28 & \\
1770 Aug 12.09311 & 06 57 05.7 & +21 37 54 & \\
1770 Aug 13.10910 & 06 59 20.7 & +21 32 43 & \\
1770 Aug 15.10289 & 07 03 59.9 & +21 23 14 & \\
1770 Aug 16.14878 & 07 06 26.5 & +21 18 06 & \\
1770 Aug 19.10084 & 07 13 50.3 & +21 03 43 & \\
1770 Aug 20.10693 & 07 16 24.9 & +20 58 18 & \\
1770 Aug 27.14605 & 07 35 16.5 & +20 17 51 & \\
1770 Aug 29.10749 & 07 40 35.2 & +20 05 09 & \\
1770 Aug 30.13372 & 07 43 25.3 & +19 59 02 & \\
1770 Aug 31.11045 & 07 46 04.3 & +19 51 29 & \\
1770 Sep 01.10342 & 07 48 48.1 & +19 45 00 & \\
1770 Sep 05.12152 & 07 59 36.1 & +19 16 25 & \\
1770 Sep 06.11063 & 08 02 15.3 & +19 08 25 & \\
1770 Sep 09.15815 & 08 10 13.7 & +18 45 37 & \\
1770 Sep 10.12304 & 08 12 45.4 & +18 37 40 & \\
1770 Sep 11.17852 & 08 15 17.4 & +18 28 51 & \\
1770 Sep 15.08826 & 08 25 10.9 & +17 58 33 & \\
1770 Sep 18.15534 & 08 32 32.8 & +17 33 54 & \\
1770 Sep 19.13952 & 08 34 52.0 & +17 25 19 & \\
1770 Sep 20.13174 & 08 37 10.3 & +17 18 47 & \\
1770 Sep 21.14191 & 08 39 27.4 & +17 10 39 & \\
1770 Sep 30.13505 & 08 58 18.1 & +15 57 48 & \\
1770 Oct 02.13471 & 09 02 06.5 & +15 42 05 & \\
1770 Oct 03.14878 & 09 04 00.8 & +15 33 40 & \\
\enddata
\end{deluxetable}

\begin{figure}
\includegraphics[width=0.5\textwidth]{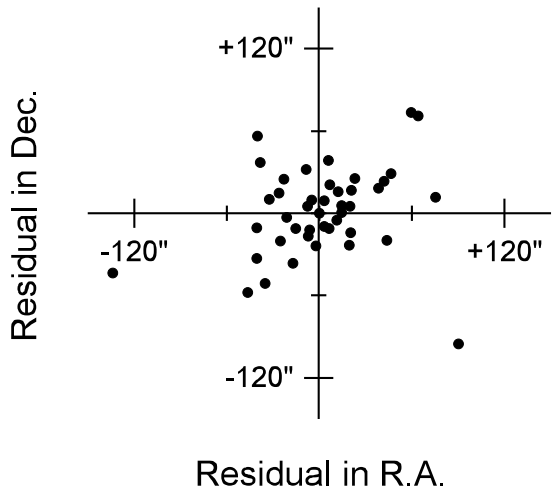}
\caption{Astrometric R.A. and Dec. residuals of Messier's observations with respect to our best fit solution in Table~\ref{tbl:orb}.\label{fig:res}}
\end{figure}

\begin{figure*}
\includegraphics[width=\textwidth]{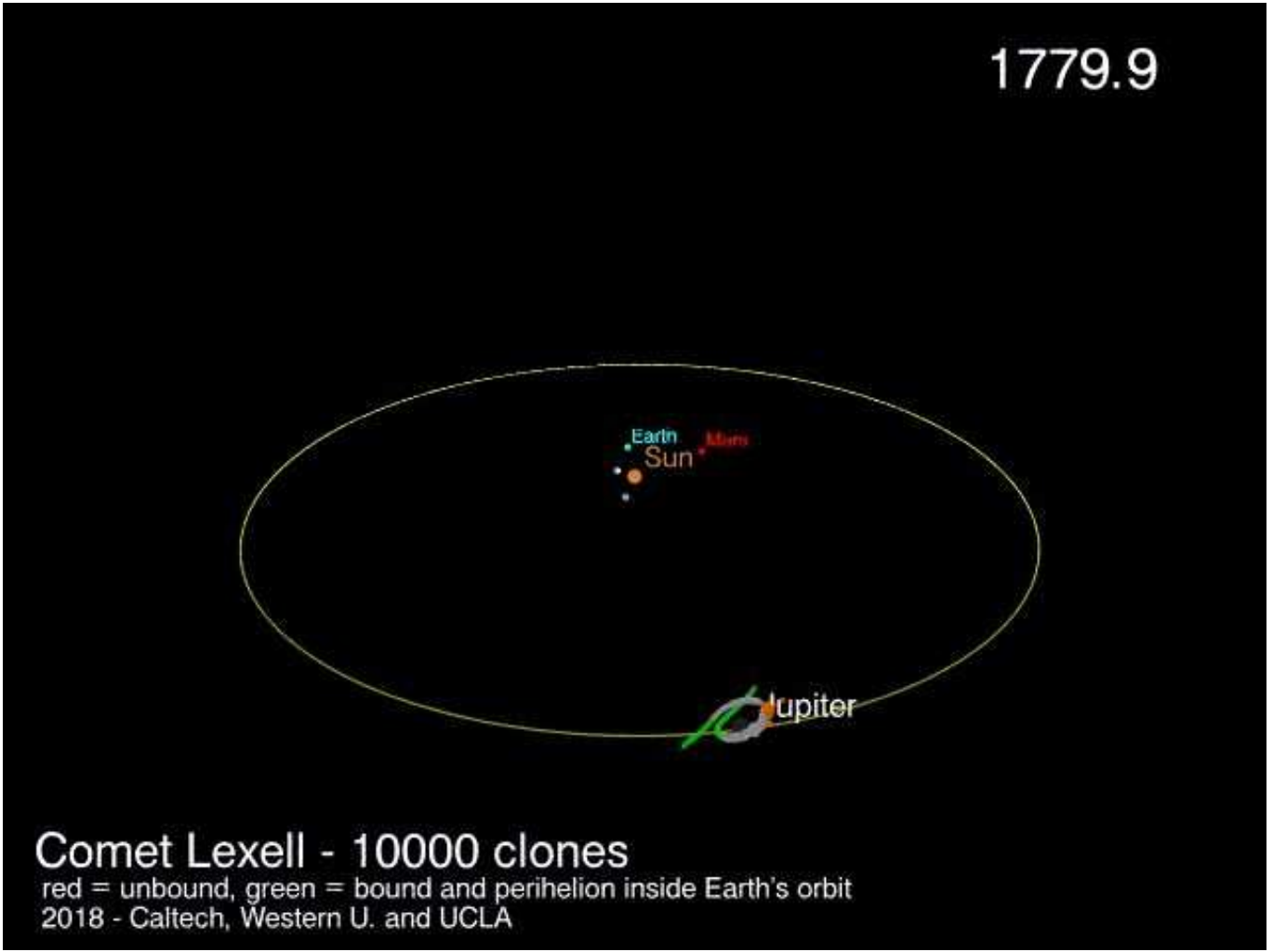}
\caption{Motions of 10~000 clones of D/Lexell from 1770 to 1790. Orange dots indicate ejected clones; grey dots indicate bound clones with perihelion outside Earth's orbit; and green dots indicate bound clones with perihelion inside Earth's orbit. It can be seen that most clones stay bound to the Solar System after the 1779 encounter with Jupiter.\label{fig:video}}
\end{figure*}

\begin{figure*}
\includegraphics[width=\textwidth]{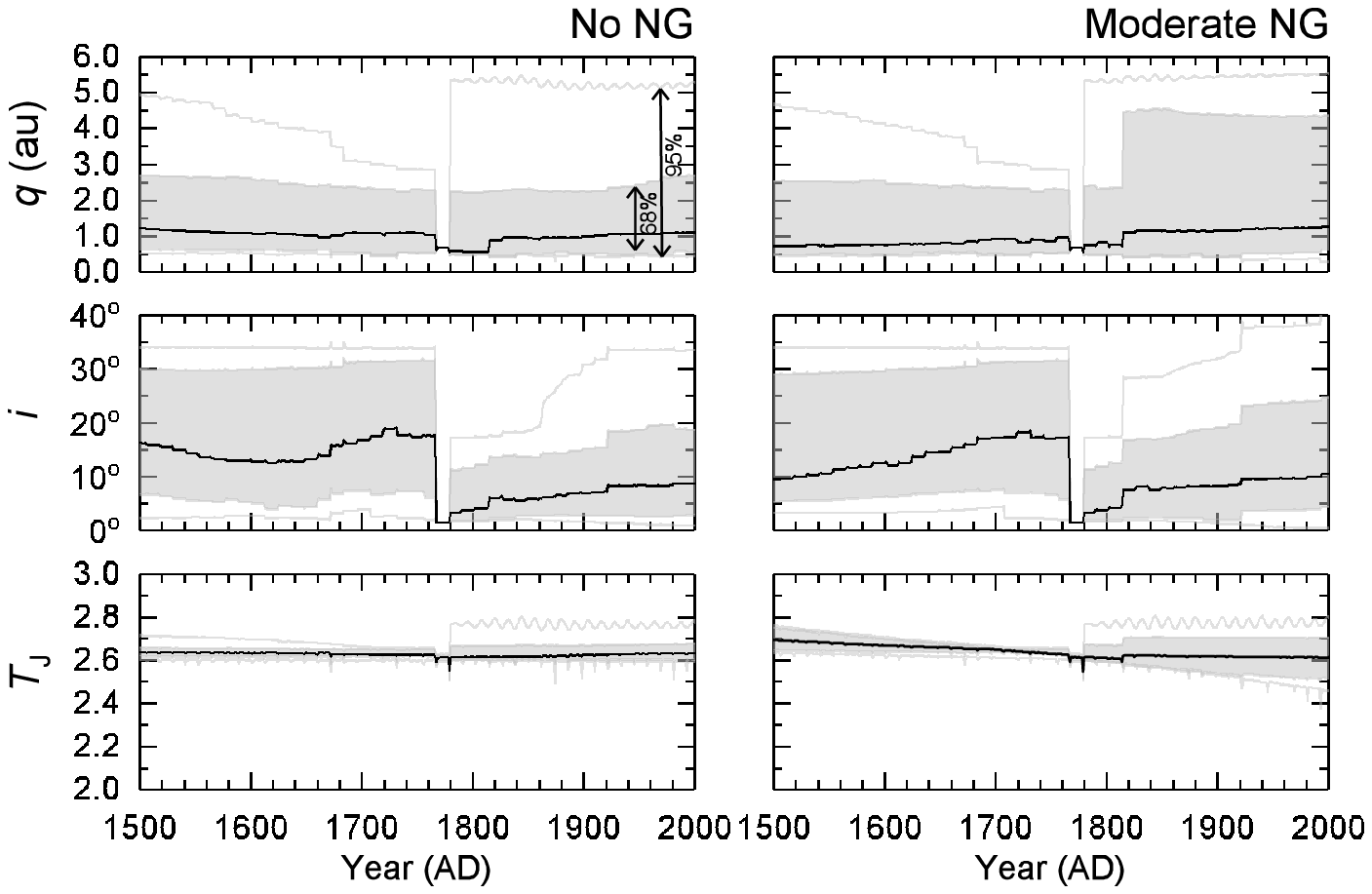}
\caption{$68\%$ and $95\%$ probability contours of D/Lexell's evolutionary path over 1500--2000~AD assuming no non-gravitational effect (left) and non-gravitational effect (right), with $\mathcal{A}_1=10^{-8}~\mathrm{au~d^{-2}}$, $\mathcal{A}_2=10^{-8}~\mathrm{au~d^{-2}}$, which is on the upper range of typical values for JFCs \citep[c.f.][]{Yeomans2004}. The black curve represents the median path.\label{fig:ev}}
\end{figure*}

The orbit of comet is calculated using the FindOrb package developed by Bill Gray\footnote{\url{https://www.projectpluto.com/find_orb.htm}.} and is tabulated in Table~\ref{tbl:orb}. Time differences between the reported observations, Terrestrial Time (TT) and Barycentric Dynamical Time (TDB) used in the numerical integrations are also being handled by FindOrb, which in turn uses the conversion table given in \citet[][p. 72]{Meeus1991}. Original notes from Messier indicated that observations on 1770 Jun 30, Jul 1, Jul 3, and Aug 7 are less accurate; these observations are excluded for our calculation. All other observations are used unweighted, assuming an astrometric precision of 1 arc-minute\footnote{This value is assigned empirically as it is not otherwise retrievable; however, considering that the angular resolution of human eye is about $1'$ \citep{Yanoff2009} and a telescope-equipped Messier must be able to achieve better resolution, the assumption of 1' is reasonable. Most of Messier's reported observations are performed with small micrometer-equipped refracting telescopes by comparing with nearby reference stars, though a few listed as 'measured without instrument' were naked-eye observations}. The root mean square (RMS) of the residuals for the best fit is 35'' (Figure~\ref{fig:res}). To determine the likely trajectory of the comet after its close approach to Jupiter, we also computed the orbit covariance, which represents the statistical orbital uncertainty as estimated from the observational data. We generate 10000 clones from the orbit covariance using a Monte Carlo scheme, and integrate them to the year of 2000 using the RADAU integrator \citep{Everhart1985}. The gravitational perturbations of the Sun, the Earth-Moon sytem with the Earth and the Moon being considered as two separate bodies, and seven other major planets are included in the force model. We find that by the year 2000, only 2.0\% of the clones had escaped or been destroyed, while 85\% remained bound to the Sun with perihelion $q < 3$~au, 40\% having $q$ within the Earth's orbit. When the same number of clones are run, each with randomly assigned non-gravitational constants of $A_1=\pm1.0$, $A_2=\pm1.0$ and $A_3=0.0$ (in units of $10^{-8}$~au~day$^{-2}$ \citep{Marsden1973}), 2.3\% are lost or destroyed by 2000, while 83\% remain bound with $q<3$~au, 45\% with $q<1$~au. Previously it was assumed that the 1779 encounter between D/Lexell and Jupiter moved it out of the inner Solar System. This encounter certainly occurred but not all of our clones suffer strong perturbations: only 1.8\% are unbound after the encounter, while 89\% remain bound with $q < 3$~au, and 68\% are bound with $q<1$~au (Figure~\ref{fig:video}). We verified this result with an independent integrator running on the Bulirsch-Stoer algorithm\citep{Bulirsch1966} whereas the Earth-Moon system is considered as a single mass. In this case 3.7\% of clones have escaped the Solar System or been destroyed by solar/planetary impacts by 2000, which is in line with the earlier result. The predominant majority of the surviving clones remain in Jupiter-family comet (JFC) like orbits (Figure~\ref{fig:ev}).

\begin{table*}
\begin{center}
\caption{Orbit of D/Lexell calculated by this work and \citet{Leverrier1844a,Leverrier1844b}, both in ecliptic J2000 reference frame. Orbital elements listed in the table are epoch, time of perihelion passage ($t_\mathrm{p}$), perihelion distance ($q$), eccentricity ($e$), inclination ($i$), longitude of the ascending node ($\Omega$), and argument of perihelion ($\omega$).\label{tbl:orb}}
\begin{tabular}{lcccccccc}
\tableline\tableline
 & Epoch (TT) & $t_\mathrm{p}$ (TT) & $q$ (au) & $e$ & $i$ & $\Omega$ & $\omega$ & Mean residual \\
\tableline
This work & 1770 Aug 14.0 & 1770 Aug 14.05 & $0.6746$ & $0.7856$ & $1.550^\circ$ & $134.50^\circ$ & $224.98^\circ$ & $35.4''$ \\
 & & $\pm0.03$~day & $\pm0.0003$ & $\pm0.0013$ & $\pm0.004^\circ$ & $\pm0.12^\circ$ & $\pm0.12^\circ$ & \\
\hline
\citet{Leverrier1844a,Leverrier1844b} & 1770 Aug 14.0\tablenotemark{a} & 1770 Aug 14.04 & $0.6744$ & $0.7861$ & $1.55^\circ$ & $134.47^\circ$ & $225.02^\circ$ & - \\
\tableline
\end{tabular}
\tablenotetext{a}{May be 1770 Aug 14.5 due to hour ambiguity.}
\end{center}
\end{table*}

Contrary to previous estimates, our statistics-based simulations argue that it is quite probable that D/Lexell remains in the inner Solar System. This argument remains valid even if we assume some non-gravitational effects such as are typically found on comets. Could D/Lexell still be wandering in the Solar System?

\section{Physical Characteristics of D/Lexell} \label{sec:pr}

To discuss the visibility of D/Lexell after 1770, we must first examine the intrinsic brightness of the comet. A highly complete compilation of brightness estimates and other morphological quantities of D/Lexell during its 1770 apparition is provided by \citet{Kronk1999b} and is tabulated in Table~\ref{tbl:mag}, with a few additional details extracted from \citet{Messier1776}. If we fit the observations with the standard formula \citep[e.g.][]{Everhart1967,Hughes1987}, $m=M_1+5\log_{10}{\varDelta}+10\log_{10}{r_\mathrm{H}}$, where $M_1$ is the absolute total magnitude of the comet, $\varDelta$ is geocentric distance and $r_\mathrm{H}$ is heliocentric distance (both in au), we find $M_1=7$. This would make D/Lexell one of the brightest comets (in terms of absolute total magnitude) that approach the Earth. 1P/Halley, for instance, has $M_1=5.5$.

\begin{table}
\begin{center}
\caption{Brightness estimates and other morphological quantities of D/Lexell reported by various observers. Extracted from \citet{Kronk1999b} unless otherwise noted.\label{tbl:mag}}
\begin{tabular}{cccccl}
\tableline\tableline
Date (1770) & Size of central condensation\tablenotemark{a} & Coma size & Tail & Magnitude & Observer \\
\tableline
Jun 14/15 & - & - & - & 5 & Messier \\
Jun 17/18 & 22'' & 5'23'' & N & - & Messier \\
Jun 22/23 & 33'' & 18' & N & - & Messier \\
Jun 24/25 & 1'15'' & 27' & - & 2 & Messier \\
Jun 27/28 & - & 0.5$^\circ$ & N & - & S. Dunn \\
Jun 29/30 & 1'22'' & 54' & - & - & Messier \\
.. & - & - & N & $<1$\tablenotemark{b} & W. Earl \\
Jul 1/2 & 1'26'' & 2$^\circ$23' & N & - & Messier \\
Aug 2/3 & 54'' & 15' & N & - & Messier \\
Aug 11/12\tablenotemark{c} & 43'' & 3'36'' & - & - & Messier \\
Aug 12/13 & - & - & N & 4--5 & Messier \\
Aug 18/19 & 38'' & - & Y & - & Messier \\
Aug 19/20 & - & - & Y & - & Messier \\
Aug 25/26\tablenotemark{c} & - & - & Y, 1$^\circ$ & - & Messier \\
Aug 27/28 & - & - & - & 5--6 & Messier \\
\tableline
\end{tabular}
\tablenotetext{a}{Called ``nucleus size'' in the original document, but by no means were the 18th century observers really observing the actual nucleus of the comet, as the nucleus would be about 1000~km in size which is highly unlikely. There is also no reason to believe that Messier and his colleagues could separate the actual nucleus from the coma using 18th century technology.}
\tablenotetext{b}{The original text was ``...larger than a star of the first magnitude'', our interpretation is that the comet was at 1st magnitude, since a star would be a point source and would not have measurable size. On the other hand, J. Six's observation on Jul. 2.0 noted ``...appeared as large as the planet Jupiter'', which we interpreted as a description of the spatial size of the cometary nucleus, since Jupiter is an extended source and its spatial size ($\sim1$') is comparable to other observations of D/Lexell near this date.}
\tablenotetext{c}{Extracted from \citet{Messier1776}.}
\end{center}
\end{table}

\begin{figure}
\includegraphics[width=0.5\textwidth]{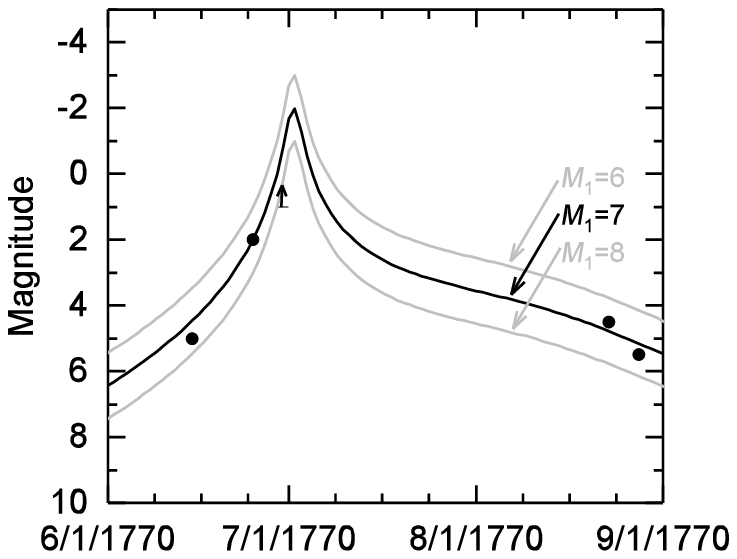}
\caption{Observed and fitted light-curve of D/Lexell during its 1770 apparition \label{fig:mag}}
\end{figure}

\begin{figure}
\includegraphics[width=0.5\textwidth]{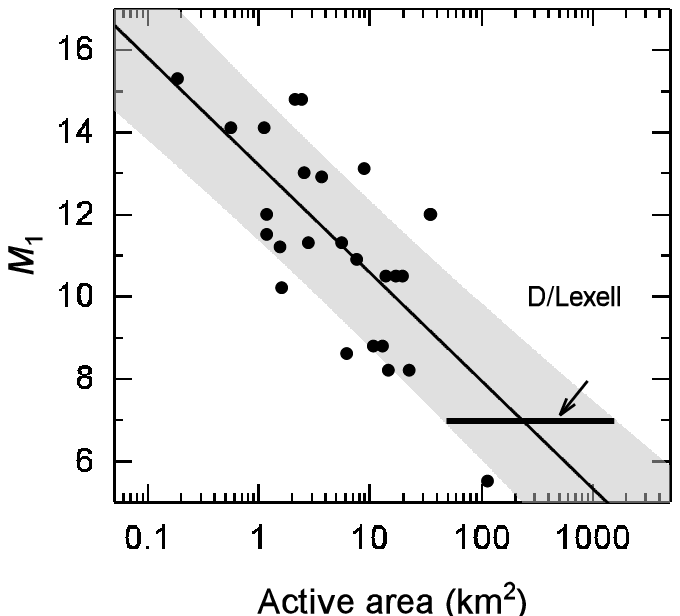}
\caption{Correlation between absolute total magnitude (a measure of the productivity of dust and gas) of the comet and size of active area on the the comet. Shaded area represents $1\sigma$ prediction level. Absolute total magnitudes are extracted from the JPL Small-Body Database (\url{https://ssd.jpl.nasa.gov/sbdb.cgi}). Cometary nucleus size and fraction of active area are extracted from \citet{Tancredi2006} except for 1P/Halley \citep{Nes1986}. Only comets with quality class QC$\leq3$ are used \citep[see the description of Table~2 in][]{Tancredi2006}. \label{fig:m1-size}}
\end{figure}

The total magnitude also provides a way to constrain the size of D/Lexell. This can be done by looking at comets whose brightness and activity have been accurately measured, such as those that have been visited by spacecraft. Based on the correlation presented in Figure~\ref{fig:m1-size}, we infer the active area of the nucleus of D/Lexell is $50$--$1600$~km$^2$. This translates to a nucleus of $4$--$22$~km in diameter taking that the fractional active area must be smaller than 1. If we take an active fraction of 0.2 \citep[which is on the high end of typical values, see][]{AHearn1995}, the corresponding diameter is $9$--$50$~km.

\begin{figure*}
\includegraphics[width=\textwidth]{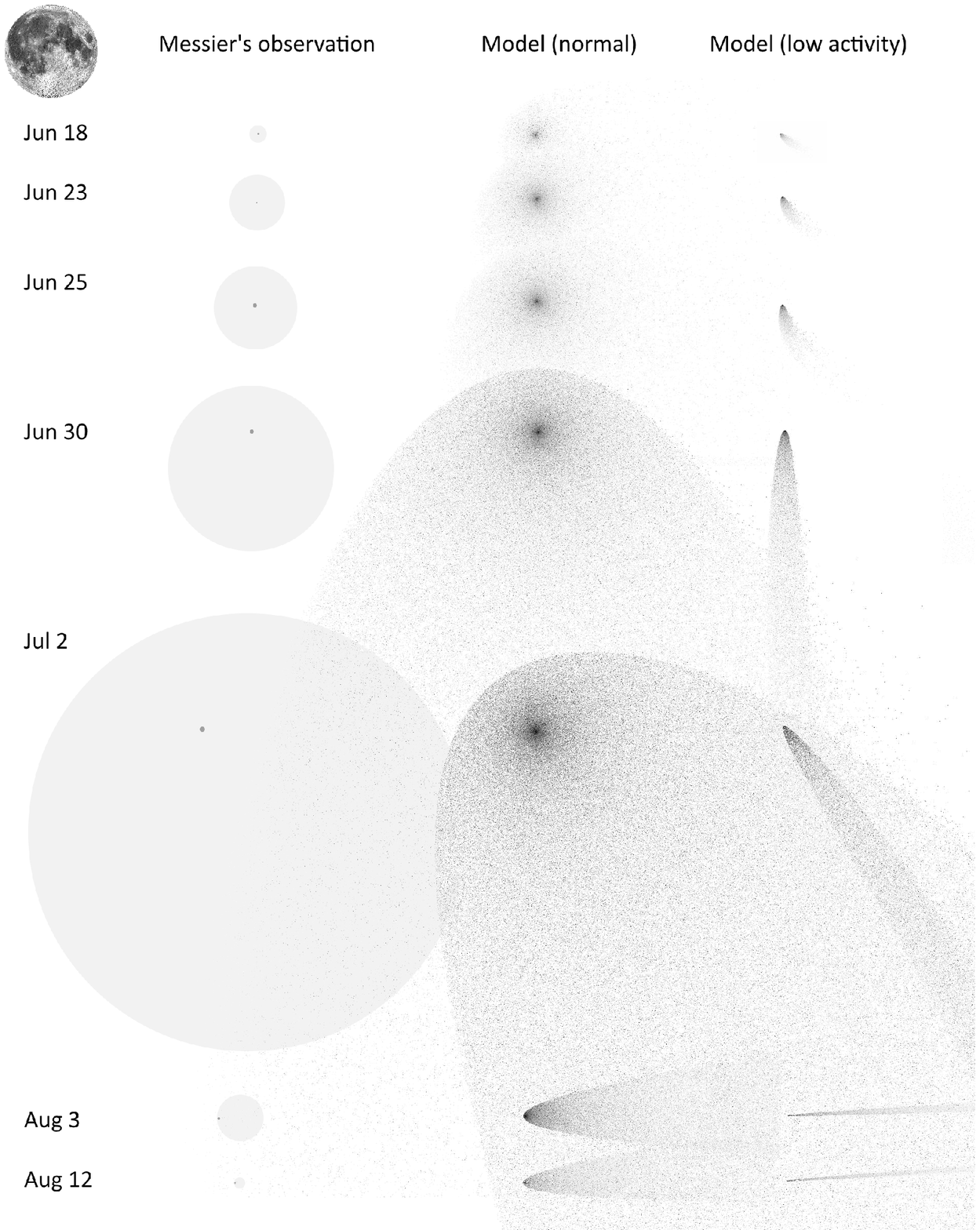}
\caption{Model images of D/Lexell created using two sets of input parameters representing normal and low levels of cometary activity as compared to Messier's observation. For comparison, the apparent size of the Moon is drawn at the upper-left corner. For the depiction of Messier's observations, filled circles in light grey represent the coma, while dots in dark grey represent the central condensation described as the ``nucleus'' by Messier (as discussed in the notes of Table~\ref{tbl:mag}). The relative position of the central condensation to the coma was not provided by Messier and is arbitrarily drawn.\label{fig:morph-mdl}}
\end{figure*}

\citet{Messier1776} also documented the apparent size of the coma and the existence of a tail in detail (summarized in Table~\ref{tbl:mag}). This enables us to model what he saw, at least at a qualitative level. Model images are created using the Monte Carlo dust code developed in \citet{Ye2014}, using two sets of input parameters representing different levels of cometary activity (Table~\ref{tbl:mdl-input}). Note that the gas component is not included in the model, as observations by Messier and others reveal a largely continuous spectrum \citep[e.g. ``silver-colored'' noted by James Six, see][]{Kronk1999b} consistent with a dominance of scattered light from dust particles. The model images, shown in Figure~\ref{fig:morph-mdl}, suggest that the activity of D/Lexell was close to average level. There is some degree of inconsistency between Messier's observations and the model images towards August 1770, where a tail is clearly seen in the model images but was not reported by Messier, despite his apparent efforts to look for one. We attribute this inconsistency to the interference from the last quarter Moon, as D/Lexell was also a morning target at that time. The tail was reported on and after Aug 18/19 as the Moon moved to the conjunction (the New Moon was on Aug 20, 1770).

\begin{table*}
\begin{center}
\caption{Input parameters for Monte Carlo simulation of coma morphology.\label{tbl:mdl-input}}
\begin{tabular}{ccccl}
\tableline\tableline
 & Normal & Low activity \\
\tableline
Dust size range ($\micron$) & 1--100 & 1--100 \\
Dust size index & -3.6 & -3.6 \\
Dust bulk density ($~\mathrm{kg~m^{-3}}$) & 2000 & 2000 \\
Dust production-heliocentric distance exponent & -4.0 & -4.0 \\
Mean ejection speed of $1~\micron$-sized dust at 1~au ($m~s^{-1}$) & 400 & 40 \\
Activate $r_\mathrm{H}$ (au) & 2.3 & 1.4 \\
Example comet & 67P/Churyumov-Gerasimenko\tablenotemark{a} & 209P/LINEAR\tablenotemark{b} \\
\tableline
\end{tabular}
\tablenotetext{a}{\citet{Ishiguro2008}.}
\tablenotetext{b}{\citet{Ye2016}.}
\end{center}
\end{table*}

\section{In Search of D/Lexell and Its Descendant} \label{sec:desc}

If D/Lexell remains in the inner Solar System and simply stopped being active, there is a good chance that it might have been recovered by modern NEO surveys as an asteroid, given the large size of the nucleus. To test this hypothesis, we examined the orbits of known asteroids and compared them to the orbit of D/Lexell. This was done by integrating the orbits of these asteroids to 1770 and computing their Southworth-Hawkins dissimilarity criterion, or the $D$-criterion \citep[][denoted as $D_\mathrm{SH}$ hereafter]{Southworth1963} against D/Lexell. The $D$-criterion is a quantitative measure of the similarity of two orbits, with smaller $D$'s indicating more similar orbits. There are several variants of the original \citet{Southworth1963} version of $D$-criterion, but since we only intend to use the $D$-criterion as a relative metric, the difference among these variants is unimportant for our purpose. Therefore we simply adopt the original expression introduced by Southworth \& Hawkins. Here we focus at NEOs of 1~km or larger as D/Lexell is likely a km-sized object as discussed above. To ensure we only examine asteroids with well-determined orbits, we focus on asteroids with orbit uncertainty number $U\leq2$ \citep{MPC1995}.

It is usually assumed that objects with $D$-criterion smaller than a certain value are likely related. However, this critical value is dependent to the size of the sample and orbital background, which varies from cases to cases. Here we start with a generous cutoff of $D_\mathrm{SH}\leq0.2$ for further examination. This particular value is loosely chosen based on the results of previous experiments \citep[c.f.][]{Drummond2000} which suggested a range of optimal cutoffs to be in the range of 0.1 to 0.2. Following the discussion in \citet{Wiegert2004} and \citet{Ye2016a}, we determine the expected number of associations with smaller $D$'s, $\langle X \rangle$. This is done in two steps: first, a set of synthetic NEO populations are generated using \citet{Greenstreet2015}'s de-biased NEO population model; second, the number of synthetic objects ($\langle X \rangle$) that have $D_\mathrm{SH}$ smaller than that of the proposed linkage is calculated.

We identify four objects that are in the proximity of D/Lexell in the $D_\mathrm{SH}$ space: 2010 JL$_{33}$ ($D_\mathrm{SH}=0.087$), 1999 XK$_{136}$ ($D_\mathrm{SH}=0.104$), 2011 LJ$_1$ ($D_\mathrm{SH}=0.171$), and 2001 YV$_3$ ($D_\mathrm{SH}=0.198$), each has a $\langle X \rangle$ of 1 in 125, 1 in 3, 6 in 1, and 14 in 1, respectively (Table~\ref{tbl:assoc}). However, the readers should bear in mind these numbers only represent the \textit{expected} number of associations one can find in a NEO population model where a large number of NEO population samples are generated; in reality, one must consider the probability of chance alignments. If we assume the local orbital distribution follows Poisson statistics, the probability of finding at least one paired object due to chance is then 

$$P(n\geq1) = e^{-\langle X \rangle} \sum_{n=1}^{\infty} \frac{\langle X \rangle ^ n}{n!}$$

\noindent Here we note that the assumption of Poisson statistics will be valid as long as the rate of object-pair due to chance is constant across the local orbital space, a region that can be understood as a quasi-infinitesimal region in the orbital space where NEO orbits are nearly uniformly distributed.

We derive the probabilities of chance alignments of 2010 JL$_{33}$, 1999 XK$_{136}$, 2011 LJ$_1$ and 2001 YV$_3$ to be $0.8\%$, $26\%$, $99.8\%$, $\sim100\%$ respectively. Apparently, 2010 JL$_{33}$ is the most promising candidate as D/Lexell's descendant. 2010 JL$_{33}$ has a diameter of 1.8~km and an albedo of 0.047, with a rotation period of 9.41~h \citep{Blaauw2011, Mainzer2011}, which is compatible to a large nucleus of D/Lexell and typical properties of cometary nuclei \citep{Snodgrass2006,Mommert2015}. By contrast, 1999 XK$_{136}$ has a smaller diameter of 0.8~km and a similarly low albedo of 0.020 \citep{Mainzer2014} with unknown rotation period. The physical properties of the less statistically significant associations, 2011 LJ$_1$ and 2001 YV$_3$, are not known.

\begin{table*}
\begin{center}
\caption{Orbit of D/Lexell (calculated by this work) compares to four possible associations: 2010 JL$_{33}$, 1999 XK$_{136}$, 2011 LJ$_1$, and 2001 YV$_3$, ordered by their $D_\mathrm{SH}$ with respect to D/Lexell. All elements are in ecliptic J2000 reference frame. Orbital elements listed in the table are epoch, time of perihelion passage ($t_\mathrm{p}$), perihelion distance ($q$), eccentricity ($e$), inclination ($i$), longitude of the ascending node ($\Omega$), and argument of perihelion ($\omega$). Also listed are the $D_\mathrm{SH}$ values with respect to D/Lexell, the expected number of km-sized NEOs with smaller $D_\mathrm{SH}$ ($\langle X \rangle$), and probability of chance alignment $P(X\geq1)$.\label{tbl:assoc}}
\begin{tabular}{lcccccccccc}
\tableline\tableline
 & Epoch (TT) & $t_\mathrm{p}$ (TT) & $q$ (au) & $e$ & $i$ & $\Omega$ & $\omega$ & $D_\mathrm{SH}$ & $\langle X \rangle$ & $P(X\geq1)$ \\
\tableline
D/Lexell & 1770 Aug 14.0 & 1770 Aug 14.05 & $0.6746$ & $0.7856$ & $1.55^\circ$ & $134.50^\circ$ & $224.98^\circ$ & - & - & - \\
2010 JL$_{33}$ & 1770 Aug 14.0 & 1770 Jun 25.32 & $0.7120$ & $0.7338$ & $4.42^\circ$ & $95.32^\circ$ & $263.38^\circ$ & $0.087$ & $0.008$ & $0.8\%$ \\
1999 XK$_{136}$ & 1770 Aug 14.0 & 1771 Aug 24.65 & $0.7003$ & $0.7073$ & $2.57^\circ$ & $71.75^\circ$ & $291.58^\circ$ & $0.104$ & $0.3$ & $26\%$ \\
2011 LJ$_1$ & 1770 Aug 14.0 & 1771 Jul 15.89 & $0.7290$ & $0.6970$ & $8.28^\circ$ & $146.86^\circ$ & $207.75^\circ$ & $0.171$ & $6$ & $99.8\%$ \\
2001 YV$_3$ & 1770 Aug 14.0 & 1771 Jan 30.41 & $0.5438$ & $0.7193$ & $5.45^\circ$ & $114.05^\circ$ & $236.71^\circ$ & $0.198$ & $14$ & $\sim100\%$ \\
\tableline
\end{tabular}
\end{center}
\end{table*}

The orbit of 2010 JL$_{33}$ is very well constrained thanks to Doppler observations taken by NASA's Goldstone Solar System Radar in 2010. Integration of 2010 JL$_{33}$ back to the year of 1770 shows little dispersal: $1\sigma$ dispersion from nominal is only $2\times10^{-5}$~au. However, with this orbit, 2010 JL$_{33}$ does not approach the Earth at the correct time to be D/Lexell. Could it be cometary non-gravitational effects that placed D/Lexell on the present orbit of 2010 JL$_{33}$?

\begin{figure*}
\includegraphics[width=\textwidth]{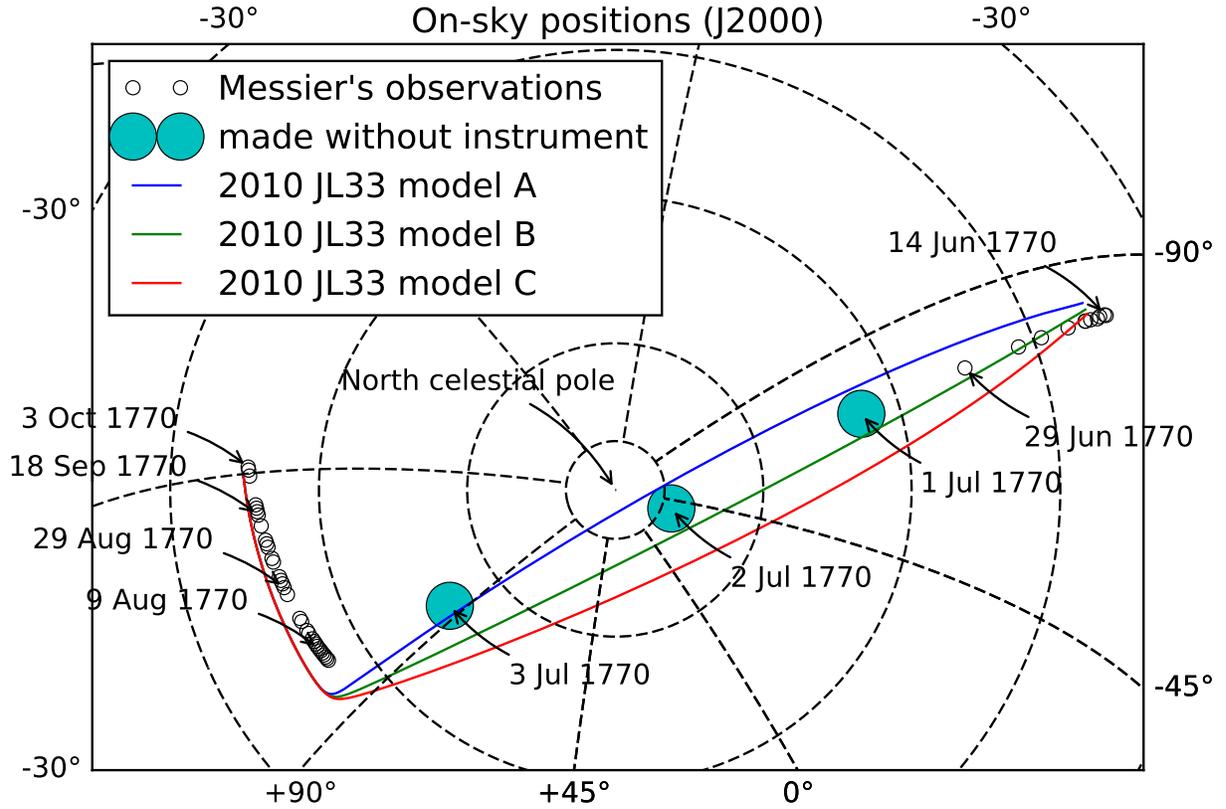}
\caption{A comparison of the on-sky paths of some of the non-gravitational models of 2010 JL$_{33}$ with the D/Lexell observations taken by Messier, most taken with the aid of a micrometeor equiped refracting telescope. Blue circles indicate those he made without instrumental aid, because of substantial cloud or in the case of those of July 3, because he was away from the observatory at a dinner with the Minister of State.\label{fig:lex-jl33}}
\end{figure*}

To test this hypothesis, we attempt to link Messier's observations of D/Lexell to modern observations of 2010 JL$_{33}$. This is first done by integrating the orbit of 2010 JL$_{33}$ backwards in time while applying some degree of non-gravitational effect, assuming that the effect remains constant over time. A wide range of parameter space is tested, covering from $\mathcal{A}_x=10^{-12}$--$10^{-6}~\mathrm{au~d^{-2}}$ \citep[where $x=1,2,3$ denotes radial, transverse and the normal directions, c.f.][]{Marsden1973} which encompass almost all known values of non-gravitational forces \citep{Yeomans2004, Hui2015}, including 10 million different combinations of $\mathcal{A}_x$. We find that a modest degree of non-gravitational effect is sufficient to bring the orbit of 2010 JL$_{33}$ into a configuration qualitatively resembling D/Lexell's 1770 passage (Figure~\ref{fig:lex-jl33}). However, getting a precise match proves much more challenging and has not yet been successfully accomplished with the constant non-gravitational model. We then consider a simple time-varying non-gravitational model: D/Lexell would initially be affected by non-gravitational effect until the epoch of $t_\mathrm{deact}$, at which point the comet deactivates and non-gravitational effect disappears. This model, however, does not yield considerably better match than the constant non-gravitational model. We also attempt to link D/Lexell to 2010 JL$_{33}$ using an orbital determination program such as FindOrb, which is also unsuccessful.

Much of the difficulty arises from the fact that the models of 2010 JL$_{33}$ experience a series of moderately close encounters with Jupiter during the $\sim200$ years in question. Consequently, the final result is extremely sensitive to the details of these encounters. This makes the search for a best fit very complicated, fraught with local minima and sharp gradients.

\section{Meteors from D/Lexell} \label{sec:meteor}

Another way to trace D/Lexell is to search for its dust footprint, detectable to the observers on the Earth as meteors. This is possible because the Earth passes close to D/Lexell's 1770 orbit twice a year, with minimum orbit intersection distance of 0.015~au in July and 0.024~au in December. The detection or non-detection of meteors from D/Lexell can be used to better constrain the orbit of the comet, as it can reveal orbital solutions of D/Lexell that are compatible or incompatible with such meteors. The presence of meteor activity can also give critical information regarding whether D/Lexell was deactivated or disintegrated at some point.

However, the chaotic nature of D/Lexell's orbit remains a major burden. As will be shown in the following, even for the period of 1767--1779 (when D/Lexell's trajectory is relatively well known), the outcome of meteor activity prediction is extremely dependent to the initial position of the parent. We approach this problem by generating 10 clones from the covariance matrix of the orbit and simulate the meteor activities from these clones. (Doing more clones becomes more computationally expensive, and we believe that 10 clones still permit a reasonable result to be derived.) The simulation uses the same setup described in \S~\ref{sec:orbit} except that radiation pressure and the Poynting-Robertson effect are now considered. We focus on the meteoroids produced by D/Lexell (and its clones) after 1767, the year that the comet was placed on its Earth-approaching orbit after a close encounter with Jupiter. Each of the 10 clones is integrated from 1770 back to 1767 as well as forward to the end point of the simulation, which we choose as the year of 2000. During the integration, each clone releases meteoroids at each of its respective perihelion passages following the meteoroid/dust ejection model described in \S~\ref{sec:pr}, except that the range of meteoroid size is from 1~mm to 10~cm and the size index is -2.8. The size of the meteoroids adopted here is slightly larger than similar studies because of the low encounter speed of the meteoroids, which means larger meteoroids are needed to produce the same amount of light; the size index of -2.8 is used as appropriated to meteoroids at such sizes \citep{McDonnell1987}. Meteoroids are integrated alongside with their parent clones, meteoroids that approach the Earth are recorded following the procedure described in \citet{Vaubaillon2005a}.

The intensity of meteor activity can be modeled from the dust production rate of the comet. Judging from the total magnitude, D/Lexell is about 10 times more active than comet 55P/Tempel-Tuttle (of which $M_1=10.0$ according to the JPL Small Body Database), the well-studied parent comet of the Leonid meteor shower, therefore we multiply the dust production rate of P/Tempel-Tuttle \citep[derived by][]{Vaubaillon2005a} by 10, which gives a dust production rate of about $10^{12}$~kg per orbit, and take it as the dust production rate of D/Lexell. The strength of meteor activity is most straightforwardly measured by the Zenith Hourly Rate (ZHR), the number of meteors per hour that an observer would see providing that the sky is dark and clear, and that the radiant is at zenith. We identify significant meteor showers that are likely to be noticed and calculate their peak time as well as the corresponding ZHR, following the technique described by \citet{Vaubaillon2005a}.

\startlongtable
\begin{deluxetable}{lclcclc}
\tablecaption{Predicted significant meteor showers from 10 clones of D/Lexell in 1770--2000. Also listed is the clone's orbit in 1770. The table is arranged by the increment of the clone's perihelion distance $q$ in 1770.\label{tbl:met}}
\tablehead{
\colhead{Clone} & \colhead{$D_\mathrm{SH}$ to nominal orbit} & \colhead{Center time (UT)} & \colhead{Duration} & \colhead{Geocentric radiant} & \colhead{Trail} & \colhead{ZHR} \\
}
\startdata
1 & 0.0016 & \multicolumn{5}{c}{$q=0.6742$~au, $e=0.7872$, $i=1.55^\circ$} \\
 && 1832 Aug 23 19:37 & 1~hr & $\alpha=266^\circ$, $\delta=+24^\circ$, $v_\mathrm{G}=13$~km/s & 1781 & 110 \\
 && 1852 Jul 13 08:32 & 3~hr & $\alpha=274^\circ$, $\delta=-16^\circ$, $v_\mathrm{G}=17$~km/s & 1770, 1776 & 50 \\
 && 1864 Aug 8 11:15 & 3~hr & $\alpha=268^\circ$, $\delta=-2^\circ$, $v_\mathrm{G}=12$~km/s & 1770, 1776 & 20 \\
 && 1887 Jul 7 19:25 & 8~hr & $\alpha=265^\circ$, $\delta=-14^\circ$, $v_\mathrm{G}=16$~km/s & 1788 & 110 \\
 && 1888 Jul 7 08:40 & 12~hr & $\alpha=265^\circ$, $\delta=-14^\circ$, $v_\mathrm{G}=16$~km/s & 1788 & 350 \\
 && 1947 Aug 21 19:26 & 6~hr & $\alpha=196^\circ$, $\delta=-59^\circ$, $v_\mathrm{G}=12$~km/s & 1851, 1858 & 30 \\
 && 1953 Aug 25 12:04 & 8~hr & $\alpha=192^\circ$, $\delta=-58^\circ$, $v_\mathrm{G}=13$~km/s & 1845, 1851 & 40 \\
 && 1958 Aug 28 17:37 & 4~hr & $\alpha=238^\circ$, $\delta=-67^\circ$, $v_\mathrm{G}=12$~km/s & 1888 & 30 \\
 && 1993 Aug 26 20:23 & 1~d & $\alpha=205^\circ$, $\delta=-50^\circ$, $v_\mathrm{G}=11$~km/s & 1864 & 90 \\
\hline
2 & 0.0012 & \multicolumn{5}{c}{$q=0.6742$~au, $e=0.7867$, $i=1.55^\circ$} \\
 && 1781 Aug 31 04:56 & 2~d & $\alpha=261^\circ$, $\delta=-11^\circ$, $v_\mathrm{G}=10$~km/s & 1770, 1776 & 120 \\
\hline
3 & 0.0012 & \multicolumn{5}{c}{$q=0.6744$~au, $e=0.7868$, $i=1.55^\circ$} \\
 && 1781 Aug 31 17:11 & 2~d & $\alpha=260^\circ$, $\delta=-10^\circ$, $v_\mathrm{G}=10$~km/s & 1770, 1776 & 270 \\
\hline
4 & 0.0011 & \multicolumn{5}{c}{$q=0.6744$~au, $e=0.7867$, $i=1.56^\circ$} \\
 && 1781 Aug 31 03:21 & 2~d & $\alpha=261^\circ$, $\delta=-10^\circ$, $v_\mathrm{G}=10$~km/s & 1770, 1776 & 90 \\
\hline
5 & 0.0010 & \multicolumn{5}{c}{$q=0.6744$~au, $e=0.7866$, $i=1.55^\circ$} \\
 && 1781 Aug 30 19:24 & 2~d & $\alpha=261^\circ$, $\delta=-10^\circ$, $v_\mathrm{G}=10$~km/s & 1770, 1776 & 40 \\
\hline
6 & 0.0005 & \multicolumn{5}{c}{$q=0.6746$~au, $e=0.7851$, $i=1.55^\circ$} \\
 && 1913 Jun 23 12:44 & 2~hr & $\alpha=277^\circ$, $\delta=-21^\circ$, $v_\mathrm{G}=26$~km/s & 1841 & 2200 \\
 && 1918 Jan 7 07:00 & 2~hr & $\alpha=285^\circ$, $\delta=-25^\circ$, $v_\mathrm{G}=26$~km/s & 1826 & 12000 \\
 && 1934 Jun 19 13:30 & 2~hr & $\alpha=275^\circ$, $\delta=-21^\circ$, $v_\mathrm{G}=28$~km/s & 1872 & 2800 \\
 && 1944 Jan 13 13:49 & 2~hr & $\alpha=289^\circ$, $\delta=-25^\circ$, $v_\mathrm{G}=27$~km/s & 1914, 1919 & 5700 \\
 && 1944 Jun 19 18:38 & 2~hr & $\alpha=275^\circ$, $\delta=-21^\circ$, $v_\mathrm{G}=27$~km/s & 1883 & 1200 \\
 && 1969 Jun 28 16:34 & 4~hr & $\alpha=280^\circ$, $\delta=-21^\circ$, $v_\mathrm{G}=26$~km/s & 1857 & 2700 \\
 && 1979 Jan 8 10:45 & 1~hr & $\alpha=287^\circ$, $\delta=-25^\circ$, $v_\mathrm{G}=26$~km/s & 1862 & 1600 \\
\hline
\multicolumn{7}{l}{Nominal orbit of D/Lexell in 1770: $q=0.6746$~au, $e=0.7856$, $i=1.55^\circ$} \\
\hline
7 & 0.0006 & \multicolumn{5}{c}{$q=0.6747$~au, $e=0.7851$, $i=1.55^\circ$} \\
 && 1894 Jan 4 20:40 & 6~hr & $\alpha=285^\circ$, $\delta=-25^\circ$, $v_\mathrm{G}=25$~km/s & 1782--1792 & 1600 \\
 && 1908 Jan 10 04:56 & 2~hr & $\alpha=286^\circ$, $\delta=-25^\circ$, $v_\mathrm{G}=27$~km/s & 1857 & 1400 \\
 && 1913 Jun 23 20:32 & 4~hr & $\alpha=277^\circ$, $\delta=-21^\circ$, $v_\mathrm{G}=26$~km/s & 1841 & 2900 \\
 && 1918 Jun 24 21:02 & 6~hr & $\alpha=277^\circ$, $\delta=-21^\circ$, $v_\mathrm{G}=26$~km/s & 1831--1836 & 1400 \\
 && 1928 Jun 21 09:40 & 4~hr & $\alpha=276^\circ$, $\delta=-21^\circ$, $v_\mathrm{G}=28$~km/s & 1826 & 5900 \\
 && 1933 Jun 21 18:29 & 6~hr & $\alpha=276^\circ$, $\delta=-21^\circ$, $v_\mathrm{G}=27$~km/s & 1826 & 10000 \\
 && 1969 Jan 5 19:17 & 1~d & $\alpha=285^\circ$, $\delta=-25^\circ$, $v_\mathrm{G}=25$~km/s & 1867, 1872 & 430 \\
 && 1970 Jul 2 02:47 & 12~hr & $\alpha=282^\circ$, $\delta=-21^\circ$, $v_\mathrm{G}=25$~km/s & 1826, 1831 & 430 \\
 && 1980 Jan 6 12:03 & 2~hr & $\alpha=285^\circ$, $\delta=-25^\circ$, $v_\mathrm{G}=25$~km/s & 1867 & 4300 \\
\hline
8 & 0.0011 & \multicolumn{5}{c}{$q=0.6749$~au, $e=0.7846$, $i=1.55^\circ$} \\
 && 1939 Jan 11 19:38 & 2~hr & $\alpha=287^\circ$, $\delta=-24^\circ$, $v_\mathrm{G}=27$~km/s & 1836 & 2400 \\
 && 1949 Jan 5 01:04 & 1~hr & $\alpha=284^\circ$, $\delta=-25^\circ$, $v_\mathrm{G}=26$~km/s & 1831 & 900 \\
 && 1949 Jun 27 02:47 & 6~hr & $\alpha=279^\circ$, $\delta=-21^\circ$, $v_\mathrm{G}=25$~km/s & 1770, 1776 & 340 \\
\hline
9 & 0.0015 & \multicolumn{5}{c}{$q=0.6749$~au, $e=0.7841$, $i=1.55^\circ$} \\
 && 1882 Jan 3 21:28 & 12~hr & $\alpha=283^\circ$, $\delta=-25^\circ$, $v_\mathrm{G}=26$~km/s & 1802--1812 & 1200 \\
 && 1913 Jul 27 01:42 & 12~hr & $\alpha=294^\circ$, $\delta=-19^\circ$, $v_\mathrm{G}=19$~km/s & 1770, 1776 & 20 \\
 && 1924 Dec 4 07:00 & 6~hr & $\alpha=267^\circ$, $\delta=-26^\circ$, $v_\mathrm{G}=19$~km/s & 1776--1812 & 550 \\
 && 1959 Jun 23 04:42 & 4~hr & $\alpha=277^\circ$, $\delta=-21^\circ$, $v_\mathrm{G}=27$~km/s & 1908 & 2500 \\
 && 1960 Jun 20 09:06 & 4~hr & $\alpha=276^\circ$, $\delta=-21^\circ$, $v_\mathrm{G}=28$~km/s & 1918 & 4400 \\
 && 1969 Jan 14 00:41 & 4~hr & $\alpha=290^\circ$, $\delta=-24^\circ$, $v_\mathrm{G}=27$~km/s & 1913 & 5800 \\
 && 1974 Jan 13 21:18 & 8~hr & $\alpha=290^\circ$, $\delta=-24^\circ$, $v_\mathrm{G}=27$~km/s & 1776--1934 & 1800 \\
 && 1979 Jan 13 23:44 & 2~hr & $\alpha=290^\circ$, $\delta=-24^\circ$, $v_\mathrm{G}=27$~km/s & 1949 & 4800 \\
 && 1983 Jul 7 06:10 & 8~hr & $\alpha=285^\circ$, $\delta=-21^\circ$, $v_\mathrm{G}=24$~km/s & 1832--1872 & 6200 \\
 && 1984 Jan 1 19:45 & 8~hr & $\alpha=283^\circ$, $\delta=-25^\circ$, $v_\mathrm{G}=24$~km/s & 1776--1872 & 11000 \\
 && 1984 Jul 1 10:36 & 2~hr & $\alpha=282^\circ$, $\delta=-21^\circ$, $v_\mathrm{G}=25$~km/s & 1908 & 2800 \\
 && 1989 Jul 7 01:56 & 8~hr & $\alpha=285^\circ$, $\delta=-20^\circ$, $v_\mathrm{G}=24$~km/s & 1832--1867 & 7200 \\
 && 1990 Jan 1 03:14 & 6~hr & $\alpha=283^\circ$, $\delta=-25^\circ$, $v_\mathrm{G}=24$~km/s & 1832--1872 & 15000 \\
\hline
10 & 0.0016 & \multicolumn{5}{c}{$q=0.6750$~au, $e=0.7840$, $i=1.54^\circ$} \\
 && 1882 Jan 4 02:49 & 8~hr & $\alpha=283^\circ$, $\delta=-25^\circ$, $v_\mathrm{G}=26$~km/s & 1807, 1812 & 900 \\
 && 1913 Dec 12 16:47 & 12~hr & $\alpha=274^\circ$, $\delta=-26^\circ$, $v_\mathrm{G}=19$~km/s & 1781--1802 & 40 \\
 && 1924 Dec 4 06:45 & 6~hr & $\alpha=267^\circ$, $\delta=-26^\circ$, $v_\mathrm{G}=19$~km/s & 1781--1807 & 600 \\
 && 1974 Jan 13 18:03 & 1~hr & $\alpha=290^\circ$, $\delta=-24^\circ$, $v_\mathrm{G}=27$~km/s & 1812 & 60 \\
 && 1983 Jul 6 04:06 & 12~hr & $\alpha=284^\circ$, $\delta=-21^\circ$, $v_\mathrm{G}=24$~km/s & 1822--1877 & 2800 \\
 && 1984 Jan 1 23:47 & 8~hr & $\alpha=283^\circ$, $\delta=-25^\circ$, $v_\mathrm{G}=24$~km/s & 1822--1872 & 7300 \\
 && 1984 Jul 1 13:41 & 6~hr & $\alpha=282^\circ$, $\delta=-21^\circ$, $v_\mathrm{G}=25$~km/s & 1887--1892 & 380 \\
 && 1989 Jul 7 05:51 & 8~hr & $\alpha=285^\circ$, $\delta=-21^\circ$, $v_\mathrm{G}=24$~km/s & 1822--1857 & 90 \\
 && 1990 Jan 1 05:06 & 6~hr & $\alpha=283^\circ$, $\delta=-25^\circ$, $v_\mathrm{G}=24$~km/s & 1822--1867 & 11000 \\
 && 1995 Jan 9 15:47 & 4~hr & $\alpha=288^\circ$, $\delta=-25^\circ$, $v_\mathrm{G}=26$~km/s & 1827--1887 & 1000 \\
\enddata
\end{deluxetable}

The result, tabulated in Table~\ref{tbl:met}, clearly shows the transition of timings and intensities of meteor showers across the orbital space of the clones. Clones with $q$ close to or larger than the nominal (clones 6--10) tend to produce stronger meteor showers, typically associated with the materials released in the 19th century; clones with $q$ smaller than the nominal (clones 1--5) produce meteor showers associated with the material released in 1770 and 1776. Since the dynamical state of D/Lexell is only relatively well known in 1767--1779, i.e. between the two close encounters to Jupiter prior to and right after the observed 1770 apparition, meteor activity from these two apparitions provides critical diagnostic information about the exact trajectory of the comet. For showers associated with the 1770--1776 ejections from clones 1--5, the radiant would have been conveniently situated in the constellation of Ophiuchus, which is easily observable in the summer months that the meteor showers are predicted to occur; for clones 2--5, meteor showers are expected to be moderately strong (as a comparison, the annual Perseid meteor shower has ZHR=100) and long-lasting thanks to the slow encounter speed and the shallow orbit of the parent. Therefore, we believe that the meteors from the 1770 and 1776 apparitions could have been noticeable even by observers of the Age of Enlightenment had D/Lexell been placed on the right orbit. On the other hand, clones 6--10 produce frequent meteor showers, with many surpassing storm level (showers with ZHR$>1000$ are defined as meteor storms) which should be easily noticeable by unaided observers.

We search for modern and historic sightings of the predicted showers. The showers in the late 19th to 20th century are relatively easy to examine due to the abundance of data; the 18th century ones are particularly challenging. One way to respond to this challenge is to consult Chinese chronicles, which have a reputation of being the most complete records for pre-modern astronomical showers. However, we note that even with Chinese records, the examination would be far from exhaustive, partially because Chinese astronomers stopped recording meteor phenomena on a regular basis after around 1650, as European missionaries introduced Aristotle's theory about meteors to royal Chinese astronomers, which was quietly accepted by the latter. Nevertheless, we diligently examine the Chinese chronicles, including the Draft History of Qing, the draft\footnote{The project was put to an end in 1930 due to the Chinese Civil War and was never completed.} of the official history of the Qing dynasty (1644--1911), Veritable Records of the Qing, as well as local, unofficial accounts. We do not find any records that match the predicted timing -- either among modern records of 20th century showers or Chinese chronicles for the 1781 event. This suggests two scenarios: either Chinese astronomers missed or did not record the 1781 event, or D/Lexell became inactive before $\sim$1800.

\begin{figure}
\includegraphics[width=0.5\textwidth]{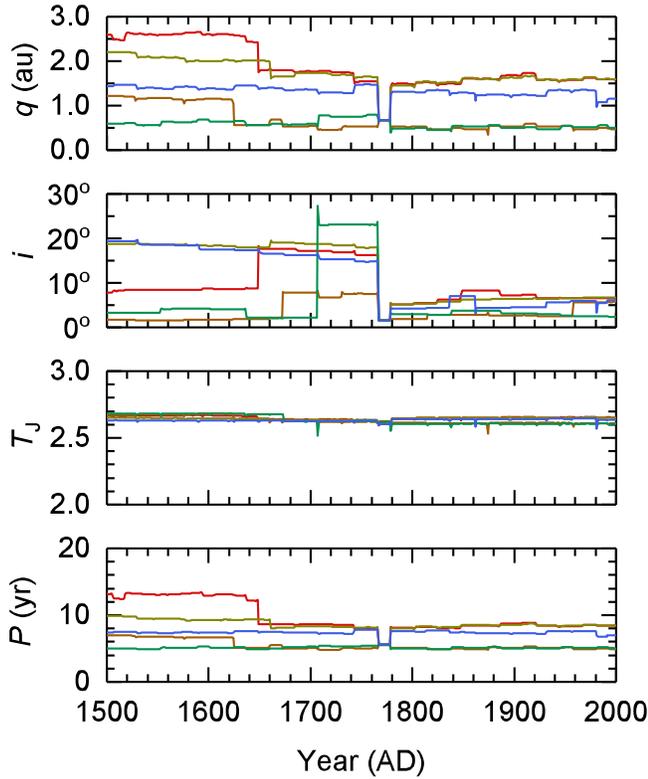}
\caption{Evolutionary paths of clones 1--5 over 1500--2000~AD assuming no non-gravitational effect.\label{fig:5clones}}
\end{figure}

Compared to historic observations, contemporary meteor surveys such as the Canadian Meteor Orbit Radar \citep{Brown2008} and Cameras for Allsky Meteor Surveillance \citep{Jenniskens2011} have much better sensitivity and consistency despite having shorter temporal coverage. To investigate meteor activity that may be detectable by contemporary and future meteor surveys, we rerun the aforementioned simulation with the same set of clones, except that the meteoroid size range is extended to 0.5~mm and the integration is continued to the year 2050. As shown in Table~\ref{tbl:met-fu}, we find that half of the clones (clones 6--10) would have produced significant meteor showers in recent years. We then search the IAU Meteor Data Center, the global clearinghouse for meteor shower detections \citep{Jopek2017}, without finding any matching records. This suggests that the true orbit of D/Lexell likely did not resemble that of any of clones 6--10. If D/Lexell was on a smaller $q$ orbit, the dynamics of the resulting meteoric materials would be much more sensitive to the details of the close approaches to Jupiter, even though the parent would likely remain in a short-period orbit (Figure~\ref{fig:5clones}). As it can be seen in Table~\ref{tbl:met-fu}, the meteor showers originating from clones with smaller $q$ (clones 0--5) have few similarities, implying that the dynamical evolution of the materials varies wildly from one clone to another. Simulation of clones on a denser grid over the orbital space will bring a clearer picture but is more computationally demanding. At this stage, we tentatively conclude that meteors potentially originating from a lower-$q$ D/Lexell will arrive in August to September from southerly radiants, and the speed will be very low.

\startlongtable
\begin{deluxetable}{lclcclc}
\tablecaption{Predicted significant meteor showers in 2000--2050 originating from the 10 clones over the apparitions of 1770 and 1776. Also listed is the clone's orbit in 1770. The table is arranged by the increment of the clone's perihelion distance $q$ in 1770.\label{tbl:met-fu}}
\tablehead{
\colhead{Clone} & \colhead{$D_\mathrm{SH}$ to nominal orbit} & \colhead{Center time (UT)} & \colhead{Duration} & \colhead{Geocentric radiant} & \colhead{Trail} & \colhead{ZHR} \\
}
\startdata
1 & 0.0016 & \multicolumn{5}{c}{$q=0.6742$~au, $e=0.7872$, $i=1.55^\circ$} \\
 && 2025 Sep 25 03:10 & 1~d & $\alpha=214^\circ$, $\delta=-53^\circ$, $v_\mathrm{G}=13$~km/s & 1770, 1776 & 5 \\
\hline
2 & 0.0012 & \multicolumn{5}{c}{$q=0.6742$~au, $e=0.7867$, $i=1.55^\circ$} \\
 && 2030 Aug 15 20:27 & 12~hr & $\alpha=254^\circ$, $\delta=+10^\circ$, $v_\mathrm{G}=11$~km/s & 1776 & 30 \\
\hline
3 & 0.0012 & \multicolumn{5}{c}{$q=0.6744$~au, $e=0.7868$, $i=1.55^\circ$} \\
\multicolumn{7}{c}{n/a} \\
\hline
4 & 0.0011 & \multicolumn{5}{c}{$q=0.6744$~au, $e=0.7867$, $i=1.56^\circ$} \\
\multicolumn{7}{c}{n/a} \\
\hline
5 & 0.0010 & \multicolumn{5}{c}{$q=0.6744$~au, $e=0.7866$, $i=1.55^\circ$} \\
 && 2043 Sep 22 14:46 & 6~hr & $\alpha=219^\circ$, $\delta=-46^\circ$, $v_\mathrm{G}=12$~km/s & 1770 & 70 \\
\hline
6 & 0.0005 & \multicolumn{5}{c}{$q=0.6746$~au, $e=0.7851$, $i=1.55^\circ$} \\
 && 2005 Jul 3 03:28 & 1~hr & $\alpha=283^\circ$, $\delta=-21^\circ$, $v_\mathrm{G}=25$~km/s & 1776 & 100 \\
 && 2026 Aug 16 15:58 & 2~hr & $\alpha=305^\circ$, $\delta=-23^\circ$, $v_\mathrm{G}=16$~km/s & 1776 & 5 \\
 && 2041 Jan 13 07:04 & 1~hr & $\alpha=290^\circ$, $\delta=-24^\circ$, $v_\mathrm{G}=27$~km/s & 1770 & 200 \\
\hline
\multicolumn{7}{l}{Nominal orbit of D/Lexell in 1770: $q=0.6746$~au, $e=0.7856$, $i=1.55^\circ$} \\
\hline
7 & 0.0006 & \multicolumn{5}{c}{$q=0.6747$~au, $e=0.7851$, $i=1.55^\circ$} \\
 && 2005 Jul 3 03:00 & 1~hr & $\alpha=283^\circ$, $\delta=-21^\circ$, $v_\mathrm{G}=25$~km/s & 1770, 1776 & 100 \\
 && 2005 Sep 24 12:38 & 3~hr & $\alpha=306^\circ$, $\delta=-22^\circ$, $v_\mathrm{G}=10$~km/s & 1776 & 10 \\
 && 2031 Jan 13 15:17 & 3~hr & $\alpha=290^\circ$, $\delta=-24^\circ$, $v_\mathrm{G}=26$~km/s & 1776 & 10 \\
 && 2036 Jan 14 04:47 & 2~hr & $\alpha=291^\circ$, $\delta=-24^\circ$, $v_\mathrm{G}=26$~km/s & 1770, 1776 & 10 \\
 && 2041 Jan 13 20:56 & 6~hr & $\alpha=291^\circ$, $\delta=-24^\circ$, $v_\mathrm{G}=27$~km/s & 1770, 1776 & 500 \\
\hline
8 & 0.0011 & \multicolumn{5}{c}{$q=0.6749$~au, $e=0.7846$, $i=1.55^\circ$} \\
 && 2005 Jan 5 19:01 & 1~d & $\alpha=286^\circ$, $\delta=-25^\circ$, $v_\mathrm{G}=25$~km/s & 1770, 1776 & 30 \\
 && 2005 Jul 3 03:21 & 1~d & $\alpha=283^\circ$, $\delta=-21^\circ$, $v_\mathrm{G}=25$~km/s & 1770, 1776 & 600 \\
 && 2016 Jan 6 17:22 & 12~hr & $\alpha=286^\circ$, $\delta=-25^\circ$, $v_\mathrm{G}=25$~km/s & 1770, 1776 & 30 \\
 && 2016 Jul 3 12:00 & 1~d & $\alpha=283^\circ$, $\delta=-21^\circ$, $v_\mathrm{G}=24$~km/s & 1770 & 10 \\
 && 2030 Jun 27 14:16 & 1~d & $\alpha=280^\circ$, $\delta=-21^\circ$, $v_\mathrm{G}=26$~km/s & 1776 & 200 \\
 && 2031 Jan 13 15:39 & 1~d & $\alpha=291^\circ$, $\delta=-24^\circ$, $v_\mathrm{G}=26$~km/s & 1770, 1776 & 50 \\
 && 2035 Jun 27 18:43 & 1~d & $\alpha=280^\circ$, $\delta=-21^\circ$, $v_\mathrm{G}=26$~km/s & 1776 & 10 \\
 && 2036 Jan 14 05:43 & 1~d & $\alpha=291^\circ$, $\delta=-24^\circ$, $v_\mathrm{G}=27$~km/s & 1770, 1776 & 300 \\
 && 2041 Jan 14 00:55 & 12~hr & $\alpha=291^\circ$, $\delta=-24^\circ$, $v_\mathrm{G}=27$~km/s & 1770, 1776 & 1600 \\
 && 2045 Jun 28 11:24 & 6~hr & $\alpha=281^\circ$, $\delta=-21^\circ$, $v_\mathrm{G}=26$~km/s & 1776 & 100 \\
 && 2046 Jun 26 06:06 & 12~hr & $\alpha=280^\circ$, $\delta=-21^\circ$, $v_\mathrm{G}=27$~km/s & 1776 & 500 \\
 && 2050 Jun 29 01:38 & 8~hr & $\alpha=281^\circ$, $\delta=-21^\circ$, $v_\mathrm{G}=26$~km/s & 1776 & 400 \\
\hline
9 & 0.0015 & \multicolumn{5}{c}{$q=0.6749$~au, $e=0.7841$, $i=1.55^\circ$} \\
 && 2005 Jan 8 21:45 & 12~hr & $\alpha=288^\circ$, $\delta=-25^\circ$, $v_\mathrm{G}=26$~km/s & 1770, 1776 & 30 \\
 && 2005 Jul 3 03:41 & 1~d & $\alpha=283^\circ$, $\delta=-21^\circ$, $v_\mathrm{G}=25$~km/s & 1770 & 100 \\
 && 2015 Jul 4 08:56 & 2~hr & $\alpha=284^\circ$, $\delta=-21^\circ$, $v_\mathrm{G}=25$~km/s & 1770, 1776 & 40 \\
 && 2016 Jan 6 14:06 & 1~d & $\alpha=286^\circ$, $\delta=-25^\circ$, $v_\mathrm{G}=25$~km/s & 1770 & 100 \\
 && 2016 Jul 2 14:49 & 1~d & $\alpha=284^\circ$, $\delta=-21^\circ$, $v_\mathrm{G}=25$~km/s & 1770, 1776 & 60 \\
 && 2025 Jun 30 12:08 & 8~hr & $\alpha=282^\circ$, $\delta=-20^\circ$, $v_\mathrm{G}=26$~km/s & 1776 & 300 \\
 && 2031 Jan 8 09:12 & 1~hr & $\alpha=288^\circ$, $\delta=-25^\circ$, $v_\mathrm{G}=25$~km/s & 1770, 1776 & 30 \\
 && 2036 Jan 15 16:41 & 1~d & $\alpha=292^\circ$, $\delta=-24^\circ$, $v_\mathrm{G}=27$~km/s & 1776 & 200 \\
 && 2050 Jun 28 17:40 & 3~hr & $\alpha=281^\circ$, $\delta=-21^\circ$, $v_\mathrm{G}=26$~km/s & 1776 & 300 \\
\hline
10 & 0.0016 & \multicolumn{5}{c}{$q=0.6750$~au, $e=0.7840$, $i=1.54^\circ$} \\
 && 2005 Jul 2 21:37 & 12~hr & $\alpha=283^\circ$, $\delta=-21^\circ$, $v_\mathrm{G}=25$~km/s & 1770 & 200 \\
 && 2016 Jan 6 01:03 & 1~d & $\alpha=286^\circ$, $\delta=-25^\circ$, $v_\mathrm{G}=25$~km/s & 1770 & 90 \\
 && 2031 Jan 8 08:14 & 1~hr & $\alpha=288^\circ$, $\delta=-25^\circ$, $v_\mathrm{G}=25$~km/s & 1770, 1776 & 10 \\
 && 2035 Jun 28 10:10 & 10~hr & $\alpha=281^\circ$, $\delta=-21^\circ$, $v_\mathrm{G}=26$~km/s & 1776 & 300 \\
 && 2036 Jan 16 01:53 & 6~hr & $\alpha=292^\circ$, $\delta=-24^\circ$, $v_\mathrm{G}=27$~km/s & 1776 & 300 \\
 && 2041 Jan 14 10:23 & 10~hr & $\alpha=291^\circ$, $\delta=-24^\circ$, $v_\mathrm{G}=27$~km/s & 1776 & 300 \\
 && 2046 Jun 26 06:05 & 15~hr & $\alpha=280^\circ$, $\delta=-21^\circ$, $v_\mathrm{G}=27$~km/s & 1776 & 400 \\
 && 2050 Jun 28 17:53 & 3~hr & $\alpha=281^\circ$, $\delta=-21^\circ$, $v_\mathrm{G}=26$~km/s & 1776 & 300 \\
\enddata
\end{deluxetable}

\section{Conclusion} \label{sec:sum}

We reviewed the case of long-lost comet D/Lexell, mainly based on a reanalysis of the observations taken by Charles Messier. We recalculated the orbit of D/Lexell and deduced the associated orbital covariance matrix, an important quantity that helped us to investigate the likely trajectory of the comet, especially after its close encounter with Jupiter in 1779. We found that there was a $98\%$ probability that D/Lexell has remained in the inner Solar System. This conclusion remains valid even if a significant degree of non-gravitational effect is considered.

From Messier's observations, we deduced that D/Lexell was one of the largest near-Earth comets currently known, with a nucleus at the order of 10~km in diameter. The activity of the comet was close to the average within the cometary population. The large size of the nucleus suggested that if D/Lexell remained in the inner Solar System, it should have been detected, either as an active comet or as an asteroid in disguise. The first scenario had been discussed throughout the 19th century and was concluded to be unlikely; we investigated the second scenario by looking among the known asteroids for orbital similarities with D/Lexell. We found asteroid 2010 JL$_{33}$ has a similar orbit to D/Lexell. A test with the NEO population model suggested that the probability of chance alignment between D/Lexell and 2010 JL$_{33}$ is 0.8\%. We unsuccessfully attempted to derive a unique orbital solution (including non-gravitational effects) that links D/Lexell and 2010 JL$_{33}$. We noted that the orbital solution was extremely dependent to the details of D/Lexell's close approaches to Jupiter, therefore the case concerning the relation between 2010 JL$_{33}$ and D/Lexell is far from conclusive.

We also simulated the dust footprint produced by a set of orbital clones of D/Lexell and found that, under certain circumstances, the footprint would be detectable at the Earth as one or more meteor showers. Clones with larger perihelion distances compared to the nominal orbit produced stronger (exceeding storm level) and more frequent meteor showers; clones with smaller perihelion distances were found to be more sensitive to close encounters with Jupiter and produced fewer meteor showers. The absence of strong meteor showers compatible with the predicted configuration suggests that the true orbit of D/Lexell resembles the latter case. This would make the dynamical evolution of associated meteoric materials more chaotic, while the parent would likely remain in a short-period orbit.

The evidence available at this stage does not allow a conclusive statement to be made. 2010 JL$_{33}$ could well be D/Lexell itself or its descendant, but establishing a dynamical pathway that places D/Lexell onto the orbit of 2010 JL$_{33}$ while satisfying every detail is challenging. Even if a pathway can be found, it would be another challenge to demonstrate that such a pathway is a unique solution of the problem rather than an ad-hoc solution that merely satisfies our assumptions. Meanwhile, careful observations of the meteors potentially originating from D/Lexell could provide important diagnostic information that would not be otherwise retrievable, which could allow post-facto orbit improvement of D/Lexell even though the comet is long lost.

\acknowledgments

We thank an anonymous referee for his/her careful reading and valuable comments that help improve the manuscript. Q.-Z. is supported by the GROWTH project (National Science Foundation Grant No. 1545949). P. W. is supported by the Natural Sciences and Engineering Research Council of Canada. M.-T. is supported by a NASA grant to David Jewitt. This work was made possible by the facilities of the Shared Hierarchical Academic Research Computing Network (SHARCNET:www.sharcnet.ca) and Compute/Calcul Canada. The authors wish to dedicate this work to Charles Messier (1730--1817), whom, besides inspiring the authors to draft this work, is most notable for publishing the celebrated Messier catalog, a compilation that has kept the authors busy during their stargazing sessions.

\end{CJK*}
\bibliographystyle{aasjournal}
\bibliography{ms}



\end{document}